# A RISC-V MCU with adaptive reverse body bias and ultra-low-power retention mode in 22 nm FD-SOI

Heiner Bauer, Marco Stolba, Stefan Scholze,
Dennis Walter, Christian Mayr
Electrical and Computer Engineering Dept., TU Dresden
Dresden, Germany
heiner.bauer@tu-dresden.de

Alexander Oefelein, Sebastian Höppner,
André Scharfe, Flo Schraut, Holger Eisenreich
Racyics GmbH
Dresden, Germany

*Abstract*— We present a low-power, energy efficient 32-bit RISC-V microprocessor unit (MCU) in 22 nm FD-SOI. It achieves ultra-low leakage, even at high temperatures, by using an adaptive reverse body biasing (ABB) aware sign-off approach, a low-power optimized physical implementation, and custom SRAM macros with retention mode. We demonstrate the robustness of the chip with measurements over the full industrial temperature range, from –40 °C to 125 °C. Our results match the state of the art (SOTA) with 4.8 uW / MHz at 50 MHz in active mode and surpass the SOTA in ultra-low-power retention mode.

## I. Introduction

IoT applications demand MCUs with low power in active as well as in sleep and retention modes. As these devices process an increasing volume of data, they also require adequate performance and large memories. We present a MCU design that addresses these conflicting requirements and is also suitable for industrial applications that require robust operation over a wide temperature range.

## II. Architecture

Our test chip architecture with a RISC-V processing element (PE) is shown in the left of Fig 1. It mirrors the PE concept in SpiNNaker 2 [1] with the aim to simplify multi-core designs. The CV32E40P processor [2] supports RV32IMXpulp instructions and is connected to four 32-KiB SRAM banks. Reverse body bias [3] is applied to the PE from an ABB generator in a zero-bias toplevel domain and enables operation at 50 MHz from 0.55 V through a dedicated supply pad.

A configurable wake-up controller is responsible for switching the PE between active and two low-power modes. Sleep mode clock gates the processor, while retention mode also puts all SRAM banks into a low-power state. Additional power can be saved by reducing the performance target of the ABB regulation during retention and operating the wake-up circuit with only 5 MHz.

## III. SRAM Retention

The single supply rail SRAM is based on the ULV dual-well 6T bitcell from [3] which allows for seamless adaptive reverse biasing together with the logic cells in the design. We enhanced the previous SRAM architecture with dedicated power state control to enable retention and power-down modes, as shown in Fig. 2. In retention mode only the SRAM periphery is powered off while the bitcell state is kept. Power-down erases all memory state but minimizes leakage power (~99 %). The wakeup procedure with in-rush current limiting takes 200 ns. The area overhead for enabling retention and power-down in a 4 KiB macro is only 2.1 %.

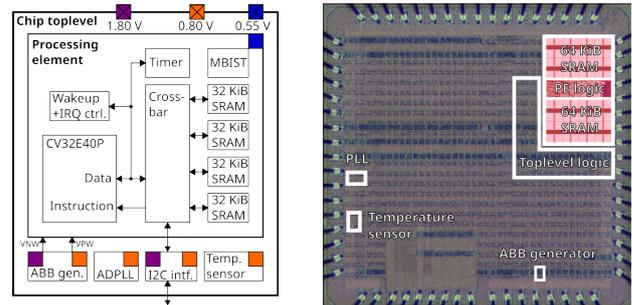

Figure 1. Block diagram (left) and annotated chip photo (right)

## IV. Implementation

Our chip (Fig. 1 right) is implemented in a 22 nm FD-SOI process from GlobalFoundries™. We used the ABB-aware methodology from [4]. Standard cells and SRAM macros have been characterized with the PVT corner dependent N-well and P-well bias voltages. Our main goal is timing robustness over the full PVT range (ssg to ffg, VDD ±10%, -40°C to 125°C) with predictable worst-case power consumption. The leakage optimized control scheme ensures that the four PVT corners slow cold, slow hot, fast cold and fast hot are bounding wrt. worst case speed and leakage power. The corner tightening benefits of ABB are fully visible to the implementation tools,

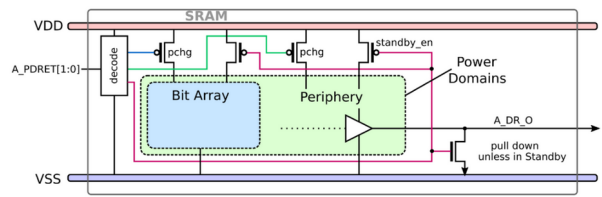

Figure 2. SRAM power state control architecture

Funded by the German Federal Ministry of Education and Research (BMBF) in the projects Scale4Edge (contract no. 16ME0136) and FastPower (contract no. 03ZZ0529A) as well as by the German Research Foundation (DFG) in project ID 390696704 – Cluster of Excellence "Centre for Tactile Internet with Human-in-the-Loop" (CeTI).

Table 1: State-of-the-art 32-bit low-power processor implementations

| | [3] | [6] | [7] | [8] | [9] | **This work** | Units |
|---|---|---|---|---|---|---|---|
| Technology | 22 nm FDSOI | 22 nm FDSOI | 28 nm FDSOI | 22 nm FDSOI | 22 nm FDSOI | 22 nm FDSOI | |
| Body bias scheme | adaptive reverse | adaptive forward | adaptive forward | static forward | static forward | adaptive reverse | |
| Processor | Arm M4 + FPU | Arm M4 + FPU | Arm M4 | RV32IMC | RV32IMCFX | RV32IMCX | |
| Architectural perf. | 3.42 | 3.42 | 3.42 | 3.20 | 3.19 | 3.19 | CM / (s MHz) |
| VDD logic / VDD SRAM | 0.55 / 0.55 | 0.50 / 0.80 | 0.40 / 0.80 | 0.65 / 0.80 | 0.50 / 0.50 | 0.55 / 0.55 | V |
| Temperature range | -40 to 125 | -40 to 125 | -40 to 85 | - | - | -40 to 125 | °C |
| Logic area | 0.043 | 0.052 | 0.145 | 0.023[a] | 0.171[a] | 0.031 | mm² |
| SRAM capacity | 256 | 84 | 64 HD, 32 ULP | 64, 32 ROM | 504, 16 SCM | 128 | KiB |
| SRAM density | 478 | 596 | 711 HD, 213 ULP | 730[a] | 650[a] | 468 | KiB/mm² |
| Frequency | 40[b] | 100[b] | 56 | 180 | 187 | 50[b] | MHz |
| Retention power at 25 | 6.6 | 46 | 7.7 | 8[f] | - | 3.2 | uW |
| Retention power at 125 | 178 | 2130 | - | - | - | 142 | uW |
| Logic PDP | 5.1 | 4.2 | 2.8 | 2.2 | - | 3.8[e] (3.1)[c,d] | uW / MHz |
| Total PDP | 6.3 | 6.9 | 4.8 | - | 6.0[e] | 4.8 (3.8)[d] | uW / MHz |
| Total energy per task | 1.8 | 2.0 | 1.4 | - | 1.9[e] | 1.5 (1.2)[d] | uJ / CM |

**a**: Estimated from chip photo **b**: Guaranteed over PVT corners **c**: De-embedded w. power analysis **d**: At 0.50 V **e**: Matmul benchmark **f**: Estimate for logic
CM: CoreMark iterations  SCM: standard cell memory  PDP: power delay product

Table 2: Power breakdown and effectiveness of SRAM bus gating

| | Logic | | SRAM | |
|---|---|---|---|---|
| | dynamic | leakage | dynamic | leakage |
| Power (no bus gating) | 308.2 uW | 2.4 uW | 69.9 uW | 4.2 uW |
| Power (with bus gating) | 211.8 uW | 2.6 uW | 51.9 uW | 4.2 uW |
| Savings vs total power | 25.1 % | -0.1 % | 4.7 % | 0 % |

which use fewer leaky cells for our performance goal of 50 MHz.

In parallel to minimizing leakage, we also wanted to reduce dynamic power while the PE is active. Therefore, we evaluated a data bus gating scheme similar to [5] and four SRAM macros sizes (1 KiB to 8 KiB). The SRAM banks in the PE are each split into smaller macros that have separate data busses driven from a central logic region. Based on the address only one macro per bank is activated and the busses to other macros in that bank are tied low. This reduces the number of concurrently active macros as well as switching power inside the buffer columns.

Based on a power analysis with CoreMark activity traces from timing annotated netlists we compare the results in Tab. 2. With bus gating implemented we save over 29% total power when the PE is active. Decreasing the SRAM macro size below 4 KiB resulted in prohibitive routing congestion and we were unable to close timing for the 50 MHz target.

V. RESULTS

We used a TP4500 thermostreamer, an on-chip temperature sensor, and a B2902A power supply with precision current sensing for our lab measurements.

The left panel of Fig. 3 shows a comparison of sleep and retention mode power over temperature. Switching from active into retention mode with a lowered ABB performance target reduces power consumption by 75x to only 3.2 uW. This surpasses all previously reported retention modes in the state of the art shown in Tab. I, especially with respect to the 128 KiB SRAM held in retention.

The power delay products (PDP) in active mode can be scaled down to 3.9 uW / MHz at the signoff border at 0.50 V, as shown on the right in Fig. 3. The PASS region was determined by PLL lock, ABB lock, MBIST pass, and successful execution of CoreMark. Backed by our robust implementation strategy, we report a performance of 50 MHz over the full PVT range, in contrast to [7]–[9], which appear to be valid only for one specific operating condition.

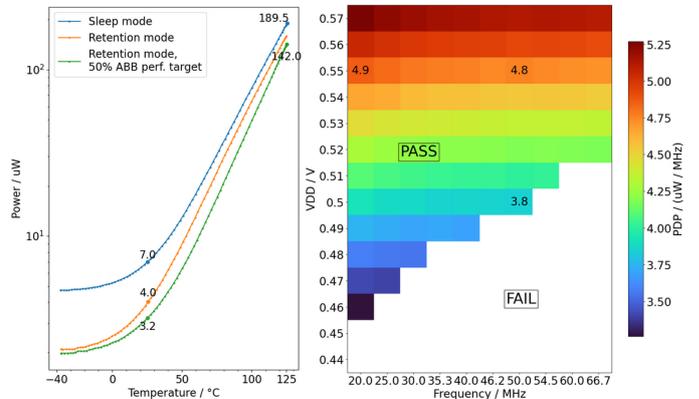

Fig. 3 Sleep and retention power (left) and CoreMark PDP shmoo (right)

VI. CONCLUSION

Our 32-bit RISC-V implementation in 22 nm FD-SOI with 128 KiB retention-enabled SRAM combines high energy efficiency in active mode with unmatched ultra-low retention power of only 3.2 uW. With a 0.5 V supply we report the lowest energy per CoreMark iteration among the state of the art. The ABB-aware implementation strategy ensures robust operation with 50 MHz over all PVT corners, including a wide temperature range up to 125 °C.

REFERENCES

[1] S. Höppner et al., arXiv:2103.08392
[2] M. Gautchi et al., TVLSI, vol. 25, no. 10, 2017
[3] D. Walter et al., IEEE Cool Chips 2020
[4] S. Höppner et al., IEEE TCAS II, vol. 67, no. 10, 2020
[5] Y. Pu et al., IEEE JSSC, vol 53, no 3, 2018
[6] S. Höppner et al., ESSDERC, 2019
[7] R. Dekimpe et al., VLSI Symposium, 2021
[8] P. Jokic et al., VLSI Symposium, 2021
[9] P.D. Schiavone et al., IEEE S3S, 2018